\documentclass[sigconf]{acmart}
\usepackage{natbib}
\usepackage[linesnumbered,ruled,vlined]{algorithm2e}
\usepackage{listings}
\usepackage{hyperref}
\usepackage{color}
\usepackage{bbding}
\usepackage{float}
\usepackage{xcolor}

\AtBeginDocument{%
  }

\acmPrice{15.00}
\acmISBN{978-1-4503-XXXX-X/18/06}
\copyrightyear{2023} 
\acmYear{2023} 
\setcopyright{acmlicensed}\acmConference[WWW '23 Companion]{Companion Proceedings of the ACM Web Conference 2023}{April 30-May 4, 2023}{Austin, TX, USA}
\acmBooktitle{Companion Proceedings of the ACM Web Conference 2023 (WWW '23 Companion), April 30-May 4, 2023, Austin, TX, USA}
\acmPrice{15.00}
\acmDOI{10.1145/3543873.3587311}
\acmISBN{978-1-4503-9419-2/23/04}
\begin{document}

\title{gDoc: Automatic Generation of Structured API Documentation}

\author{Shujun Wang}
\email{wangshujun.wsj@alibaba-inc.com}
\orcid{0000-0001-8161-1695}
\affiliation{%
 \institution{Alibaba Group}
 \city{Beijing}
 \country{China}
 \postcode{100000}
}

	\author{Yongqiang Tian}
\email{puling.tyq@taobao.com}
\affiliation{%
	\institution{Alibaba Group}
	\city{HangZhou}
	\country{China}
	\postcode{100000}
}

\author{Dengcheng He}
\email{dengcheng.hedc@alibaba-inc.com}
\affiliation{%
	\institution{Alibaba Group}
	\city{HangZhou}
	\country{China}
	\postcode{100000}
}

\renewcommand{\shortauthors}{Wang et al.}


\begin{abstract}
	Generating and maintaining API documentation with integrity and consistency can be time-consuming and expensive for evolving APIs. To solve this problem, several approaches have been proposed to automatically generate high-quality API documentation based on a combination of knowledge from different web sources. However, current researches are weak in handling unpopular APIs and cannot generate structured API documentation. Hence, in this poster, we propose a hybrid technique(namely \textit{gDoc}) for the automatic generation of structured API documentation. We first present a fine-grained search-based strategy to generate the description for partial API parameters via computing the relevance between various APIs, ensuring the consistency of API documentation. Then, we employ the cross-modal pretraining Seq2Seq model M6 to generate a structured API document for each API, which treats the document generation problem as a translation problem. Finally, we propose a heuristic algorithm to extract practical parameter examples from API request logs. The experiments evaluated on the online system show that this work's approach significantly improves the effectiveness and efficiency of API document generation.
\end{abstract}


\maketitle

\section{Introduction}
Application Programming Interfaces (APIs) play an essential role in modern software development\cite{API,rec13,rec26}. With the help of APIs, developers can complete their tasks more efficiently\cite{APIEffecient,apirec_rack}. OpenAPI indicates a behavior where producers offer Application Programming Interfaces (APIs) to help end-users access their data, resources, and services. Increasingly, many companies have published an ocean of OpenAPIs on the internet\cite{oceanAPI}. For example, as of January 2023, Alibaba Cloud lists over 14500 OpenAPIs for web services(https://next.api.alibabacloud.com/home). 

When learning to use an Application Programming Interface (API), programmers need to understand the API functions' inputs and outputs (I/O) by studying API documents. Table~\ref{tab:doc} exhibits a fragment of structured API documentation. API documentation typically contains multiple parameters, each with a description and an example—unfortunately, numerous parameters cause the generation, update, and maintenance of API documentation to be time-consuming. 

\begin{figure}[h]
	\centering
	\includegraphics[width=0.9\linewidth]{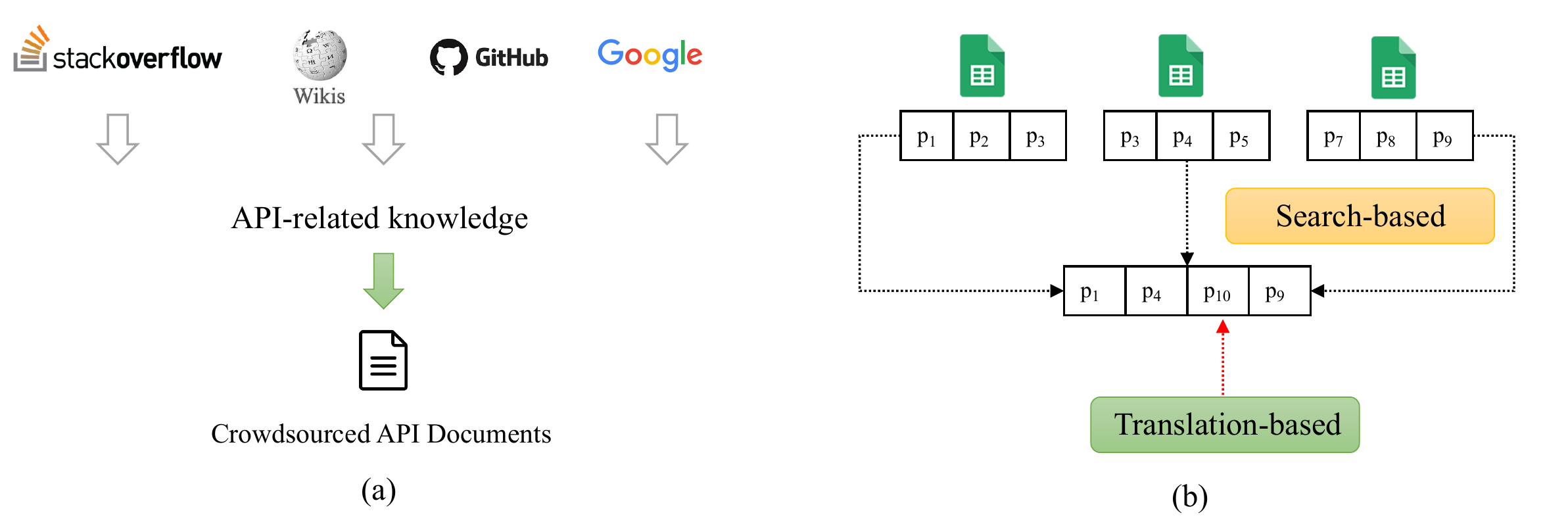}
	\caption{Existing Work Vs. gDoc}
	\label{contribution}
\end{figure}

To ease this problem, in recent years, many data-driven methods for automatic generation of unstructured API documentation have been proposed\cite{apidoc_overview,chen2014asked,autogdoc_peng}. These methods mine and integrate scattered API-related information on the web to serve as API documentation(See Figure~\ref{contribution}(a)). Unfortunately, previous approaches suffer from three significant drawbacks:
\begin{itemize}
	\item 	 API-related information collected from the web cannot be organized in a uniform and structured form(Such as Table~\ref{tab:doc}).
	\item  	 Previous methods are weak in handling unpopular APIs because API-related information on the web usually revolves around some hot APIs.
	\item  	 It is difficult for existing methods to guarantee content consistency at parameter granularity, such as when two APIs have the same parameters.
\end{itemize}

\begin{table}
	\caption{A Fragment of API Documenration\label{tab:doc}}
	\begin{tabular}{|c|c|c|c|c|} 
		\hline
		Parameter & Type & Required & Description & Example\\
		
		\hline 
		\scriptsize TemplateCode & \scriptsize String & \scriptsize Yes & \scriptsize Message Template ID&\scriptsize SMS\_123456 \\ 
		\hline
		\scriptsize PhoneNumber & \scriptsize String & \scriptsize Yes & \scriptsize Phone Number &\scriptsize 186****9602 \\
		\hline
	\end{tabular}
\end{table}

In this poster, we develop \textit{gDoc}, an automated structured API document generation approach. The contributions of gDoc are outlined as follows:
\begin{itemize}
	\item gDoc regards API documentation as a structured document composed of multiple parameters, and based on this, it mines the internal relationship between APIs to realize parameter-level API documentation generation(See Figure~\ref{contribution}(b) black dotted line).
	\item gDoc adopt the Seq2Seq model to translate API meta to API documentation, treating the document generation problem as a translation problem. Hence, gDoc can generate a document for any API(See Figure~\ref{contribution}(b) red dotted line).
	\item We propose a heuristic algorithm to extract valuable parameter examples from API request logs. 
	\item gDoc has been fully applied to the generation of Alibaba Cloud API documents. The adoption rate of parameter descriptions and examples generated by gDoc exceeds 80\%.
\end{itemize}

\section{Running Example}
\begin{figure}[!h]
	\centering
	\includegraphics[width=0.9\linewidth]{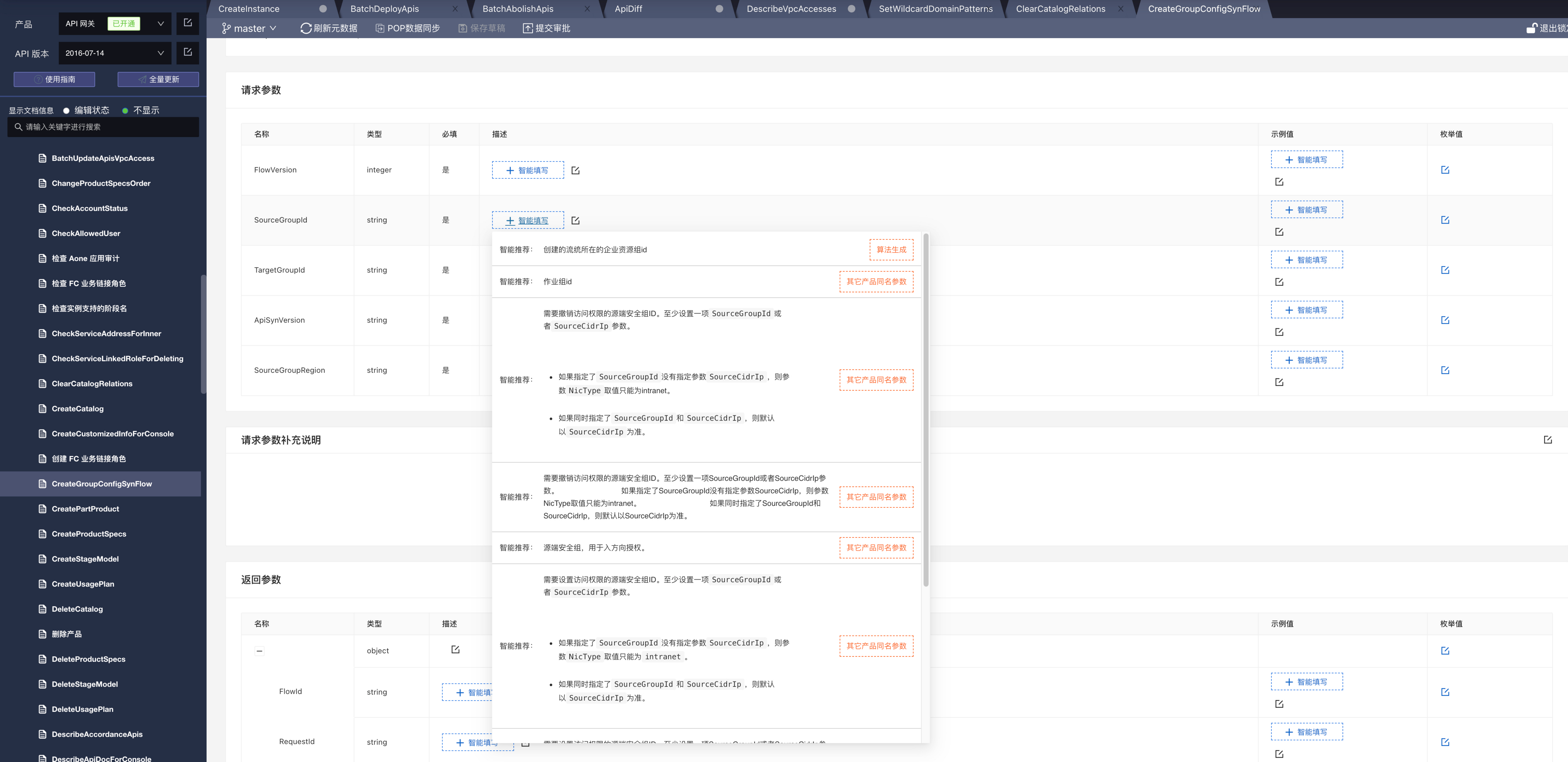}
	\caption{Demonstration of API Documentation Generation}
	\label{recommend}
\end{figure}

As shown in Figure~\ref{recommend}, we demonstrate an automated production process for API documentation. When the documentation engineer clicks into an API document, the blank fields in the document will be assigned a series of recommended values (hidden under the smart recommendation button).
\section{Search-based Generation}
\textbf{A Phenomenon:} we have noticed that the same parameter may be widely present in multiple API documents.
\begin{table}[htbp]
	\begin{minipage}[t]{0.45\linewidth}
		\caption{SendSms\label{tab:sendsms}}
		\centering
		\begin{tabular}{c|c} 
			\hline
			\small	Input  & \small Output \\
			\hline 
			\scriptsize PhoneNumbers & \scriptsize Code \\
			\scriptsize \color{red}{SignName} & \scriptsize Message  \\
			\scriptsize TemplateCode &\scriptsize BizId \\
			\scriptsize TemplateParam &\scriptsize  RequestId \\
			\scriptsize OutId &\\
			\hline
		\end{tabular}
	\end{minipage}
	\begin{minipage}[t]{0.45\linewidth}
		\caption{ AddSmsSign\label{addsmssign}}
		\centering
		\begin{tabular}{c|c} 
			\hline
			\small Input  & \small Output \\
			\hline 
			\scriptsize \color{red}{SignName} & \scriptsize Code \\
			\scriptsize Remark & \scriptsize Message  \\
			\scriptsize SignFileList & \scriptsize \color{red}{SignName} \\
			\scriptsize TemplateParam & \scriptsize RequestId \\
			\hline
		\end{tabular}
	\end{minipage}
\end{table}
Table~\ref{tab:sendsms}, \ref{addsmssign} exhibits the input parameters and output parameters of two APIs(i.e., SendSms, AddSmsSign). 

\textbf{Assumption :} we believe that the presence of parameters in multiple API documents is widespread. Furthermore, we assume that the same parameter may have the same semantics in different API documents (i.e., with the same description and example).

\begin{figure}[htpb]
	\centering
	\includegraphics[width=0.5\linewidth]{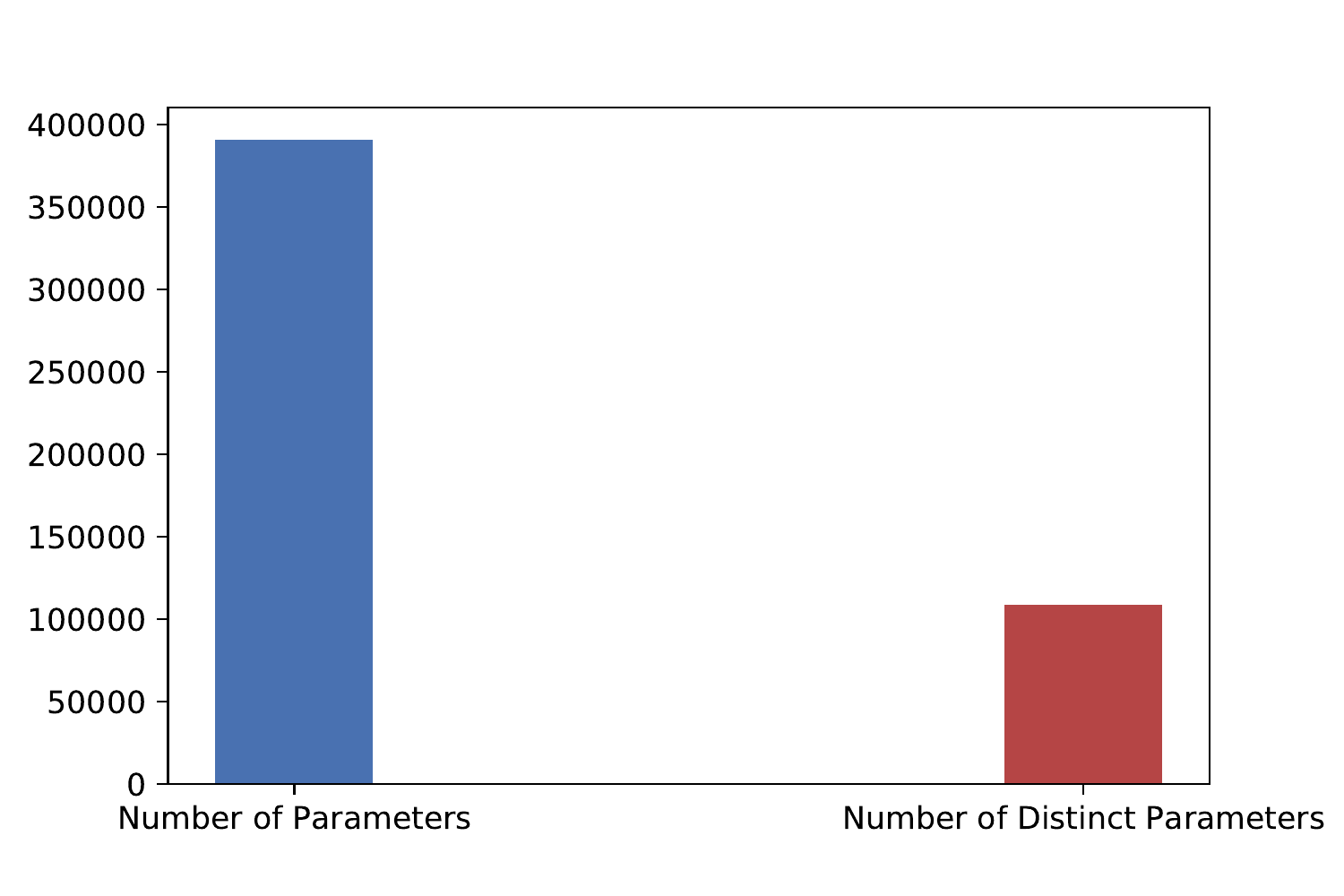}
	\caption{Comparison of the Number of Parameters}
	\label{shrink}
\end{figure}

\textbf{Analysis:} 
As shown in Figure~\ref{shrink}, we conducted a fine-grained analysis of Alibaba Cloud's 29217 APIs. 
These APIs have a total of 390796 parameters. After deduplication, the total number of parameters is 108882, and the compression ratio is 3.56, which means that a parameter will appear 3.56 times in various API documents on average.
In Figure~\ref{shrink}, it is assumed that all the same parameters have the same meaning, but this Assumption may not necessarily hold; thus, we further verify this Assumption. We cluster the parameter descriptions according to the parameter names and take the similarity of the two most similar descriptions in the cluster as the set similarity. In the end, we found that the average similarity of all sets was 0.72, which proved the correctness of our hypothesis. That is to say, parameters with the same name are more likely to have the same meaning.

\textbf{Accomplish:} Given a blank API document $D_i$, for any parameter named $p_{i,j}$ in $D_i$, we recommend content named $p_{i,j}$ with description and example values in other API documents as candidate recommendation item. This method can improve the production efficiency of API documents and improve the consistency of API documents.
\section{Translation-based Generation}
Search-based generation does not guarantee that candidate descriptions will be generated for all API parameters. Therefore, in this section we propose a translation-based approach to produce a high-quality description for all parameters.

Natural language generation (NLG), also known as text generation, is one of the essential tasks in natural language processing\cite{wang2020answering}. This paper concentrates on applying the Seq2Seq model to solve the OpenAPI documentation generation problem, dramatically reducing document production and maintenance overhead.

\begin{table}[h]
	\caption{OpenAPI Documentation\label{tab:apigeneration}}
	\begin{tabular}{l|c|c|l|c} 
		\hline
		\small	\textbf{Parameter} &\small \textbf{Type} &\small \textbf{Required} &\small \textbf{Description} &\small \textbf{Example} \\
		\hline 
		\scriptsize SignName & \scriptsize String & \scriptsize Yes & \scriptsize Signature Name& \scriptsize Aliyun\\
		\scriptsize PhoneNumbers &\scriptsize String &\scriptsize Yes&\scriptsize Phone Numbers&\scriptsize [186****9602] \\
		\scriptsize OutId &\scriptsize String &\scriptsize No &\scriptsize Extension Field&\scriptsize abc \\
		\hline
	\end{tabular}
\end{table}

Table~\ref{tab:apigeneration} exhibits an example of OpenAPI documentation. In our work, we mainly focus on the \textit{Description} generation by translating the combined OpenAPI parameter name and OpenAPI name.

We first introduce the RNN-based Seq2Seq model. A recurrent neural network (RNN) model recurrently calculates a vector named recurrent state or hidden state $h_{n}$ by taking a sequence of words $\left\{\omega_{1}, \omega_{2}, \ldots, \omega_{N}\right\}:$
$$
h_{n}=f\left(h_{n-1}, \omega_{n}\right), n \in(1, N), h_{0}=0
$$
The $h_{0}$ denotes the initial state and is always set as zero at the training time. Usually, $h_{n}$ depends on the current word $\omega_{n}$ and previous ones before the current time step. Equation $1, f$ denotes a parametrized nonlinear function: sigmoid, hyperbolic tangent, long-short term memory (LSTM), and recurrent gate unit (GRU). The hidden state will lose long contextual information when a vanilla RNN, such as a sigmoid or hyperbolic tangent, is used. LSTM or GRU can handle longer-term contexts by bringing in a memory cell. Moreover, GRU requires less computational cost compared with LSTM. Thus, GRU is used as the RNN cell unit. The equations of GRU are summarized as follows:
$$
\begin{gathered}
	z_{t}=\sigma\left(W_{z} \omega_{t}+U_{z} h_{t-1}\right) \\
	r_{t}=\sigma\left(W_{r} \omega_{t}+U_{r} h_{t-1}\right) \\
	\widetilde{\mathrm{h}}_{\mathrm{t}}=\tanh \left(\mathrm{W} \omega_{\mathrm{t}}+\mathrm{U}\left(\mathrm{r}_{\mathrm{t}} * \mathrm{~h}_{\mathrm{t}-1}\right)\right) \\
	h_{t}=\left(1-z_{t}\right) * h_{t-1}+z_{t} * \tilde{h}_{t}
\end{gathered}
$$
In the Equation above, the $\sigma$ is the nonlinear function, i.e., logistic sigmoid, which limits output to range $[0,1]$. $z_{t}$ is the update gate deciding the weight of input information past, and $r_{t}$ is the reset gate determining the weight of the last state. The candidate updates $\tilde{h}_{t}$ and controls the percentage of information obtained from $h_{t-1}$ with a reset gate. The final update $h_{t}$ depends on the update gate and candidate update. The subscript letter $t$ represents the time step.

In this paper, we employ the Seq2Seq model proposed by M6\cite{m6}, a cross-modal pretraining method, referring to Multi-Modality to MultiModality Multitask Mega-transformer, for unified pretraining on the data of single modality and multiple modalities. M6 scales the model size to 10 billion and 100 billion parameters and builds the Chinese's largest pre-trained model. This paper applies the M6 model to OpenAPI documentation generation applications and demonstrates its outstanding performance.

\section{Parameter Example Generation}
The examples of the parameter are essential. An intuitive way to obtain examples of parameters is to extract them from the API request log. However, randomly extracting examples from massive values is not enough. Take Alibaba Cloud API request logs as an example. There are more than 16 billion OpenAPI requests every day. Analyzing the total amount of logs every day requires strong hardware resource support. More importantly, it will waste much time. We believe that the example values of parameters should be universal. For example, 90\% parameter values are composed of pure English letters, and the examples should be one of the values of this 90\% rather than others.

\IncMargin{1em}
\begin{algorithm} \SetKwData{Left}{left}\SetKwData{This}{this}\SetKwData{Up}{up} \SetKwFunction{Union}{Union}
	\SetKwFunction{ExtractCommonSubsequence}{ExtractCommonSubsequence}
	\SetKwFunction{TransformAndCompress}{TransformAndCompress}
	\SetKwFunction{LengthComputing}{LengthComputing}
	\SetKwFunction{MergeSubsequence}{MergeSubsequence}
	\SetKwFunction{MergePatterns}{MergePatterns}
	\SetKwFunction{MergeLengths}{MergeLengths}
	\SetKwInOut{Input}{Input}\SetKwInOut{Output}{Output}
	
	\Input{A set of parameter values $V = v_1 \cup v_2 \cup \ldots \cup v_n$} 
	\Output{A common subsequence $s$ \\ A set of parameter patterns $P = p_1 \cup \ldots \cup p_n  $ \\ A set of parameter lengths $L = l_1 \cup l_2 \cup \ldots \cup l_n$}
	\BlankLine 
	
	\textbf{Mapper}
	
	\qquad $s_i \leftarrow$\ExtractCommonSubsequence{$v_i$}\; 
	\qquad $p_i \leftarrow$\TransformAndCompress{$v_i$}\; 
	\qquad $l_i \leftarrow$\LengthComputing{$v_i$}\; 
	
	\textbf{Reducer}
	
	\qquad $s \leftarrow$\MergeSubsequence{$s_i$}\;
	\qquad $P \leftarrow$\MergePatterns{$p_i$}\;
	\qquad $L \leftarrow$\MergeLengths{$l_i$}\;
	\caption{Parameter Abstraction}
	\label{metaparam} 
	\Return $s$, $P$, $L$
\end{algorithm}
\BlankLine
\DecMargin{1em} 

As shown in Algorithm~\ref{metaparam}, we take a two-stage approach: Mapper and Reducer, to extract common parameter features from an ocean of values. Specifically, we emphasize three types of parameter features:
\begin{itemize}
	\item Component Element Analysis.
	\item Element Arrangement Analysis.
	\item Length Analysis.
\end{itemize}

There are two main stages of parameter abstraction:

\textbf{Mapper:} In the mapper stage, we mainly focus on extracting local features of parameter values. Specifically,
\begin{itemize}
	\item \textbf{Extracting Common Subsequence} will extract the longest common subsequences in the parameter set $v_i$.
	\item \textbf{Transformer And Compress} will translate a specific parameter value into an abstract value. The translation rules are as follows:
	\begin{table}[!h]
		\caption{Transform Rules\label{tab:transform}}
		\begin{tabular}{l|l} 
			\hline
			Representation & Description \\
			\hline  
			\textit{z} & The Chinese character \\
			\textit{x} & English lowercase characters  \\
			\textit{X} & English uppercase characters \\
			\textit{d} & Number  \\
			& Other characters reserved\\
			\hline
		\end{tabular}
	\end{table}
	The transformer stage is responsible for abstract parameters. For example, $123$ will be abstracted as $ddd$, and the compressing stage will further translate it into a character $d$.
	\item \textbf{LengthComputing} mainly computes the length information of parameters and their frequency.
\end{itemize}
\textbf{Reducer:} In this stage, the local parameter features are combined to extract the global features of the parameters.

An example output of algorithm~\ref{metaparam} is as follows:
\begin{lstlisting}[language = C++, numbers=left, 
	numberstyle=\tiny,keywordstyle=\color{blue!70},
	commentstyle=\color{red!50!green!50!blue!50},frame=shadowbox,
	rulesepcolor=\color{red!20!green!20!blue!20},basicstyle=\ttfamily]
{
 "parameter_pattern": "X_d",
 "rate": 0.994,
 "examples": [
  "SMS_41515455",
  "SMS_210775383",
  "SMS_216370632",
  "SMS_173474376",
  "SMS_186610582",
  "SMS_182400031"
 ],
 "common_100": "",    Common string for 100%
 "common_80": "SMS_", Common string for 80% 
 "common_60": "SMS_"  Common string for 60%
}
\end{lstlisting}
\begin{figure*}[!h]
	\centering
	\begin{minipage}[t]{0.32\linewidth}
		\centering
		\includegraphics[width=0.88\linewidth]{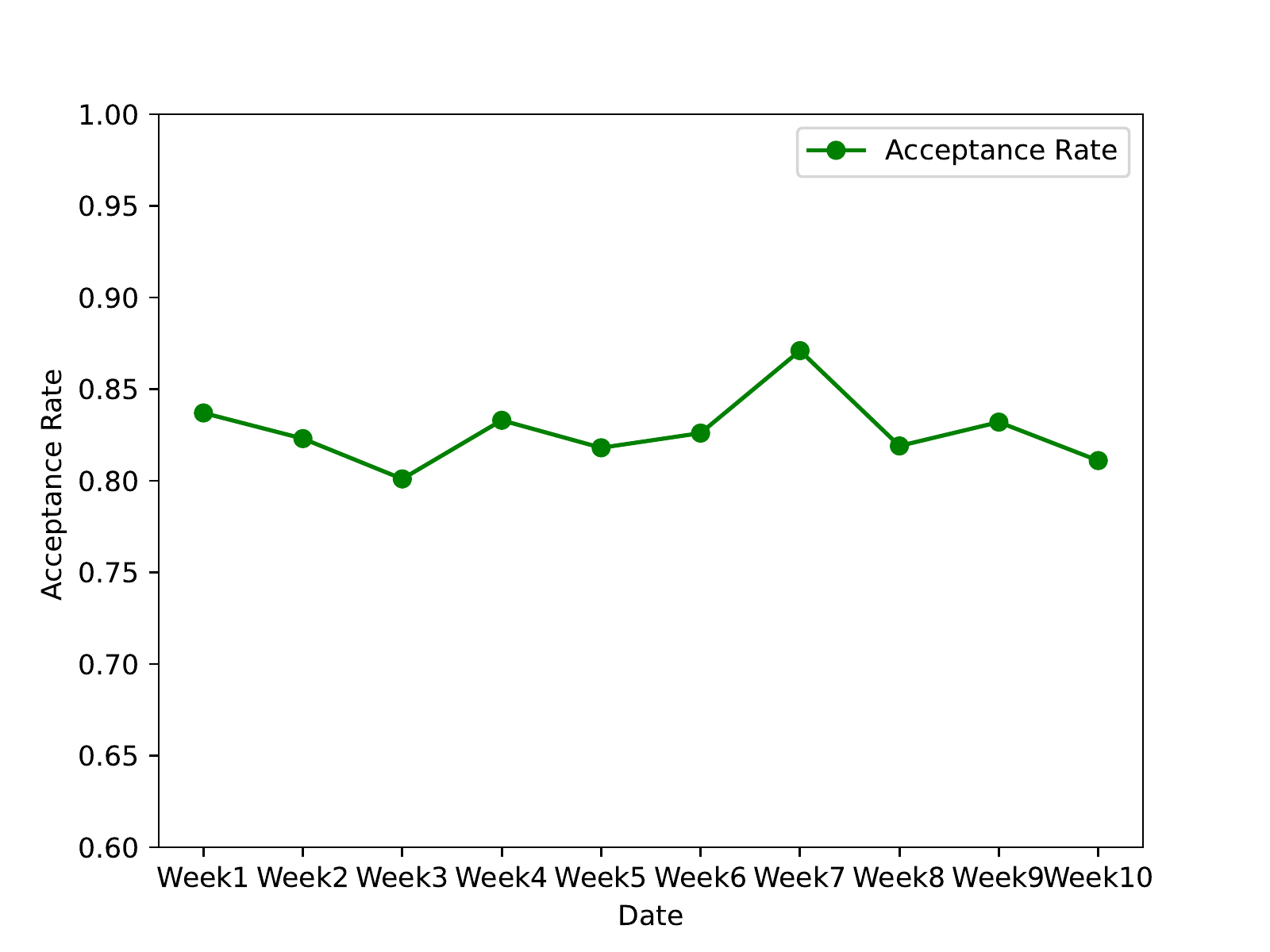}
		\caption{\small Overall Evaluation}
		\label{overall_acceptance_rate}
	\end{minipage}
	\begin{minipage}[t]{0.32\linewidth}
		\centering
		\includegraphics[width=0.88\linewidth]{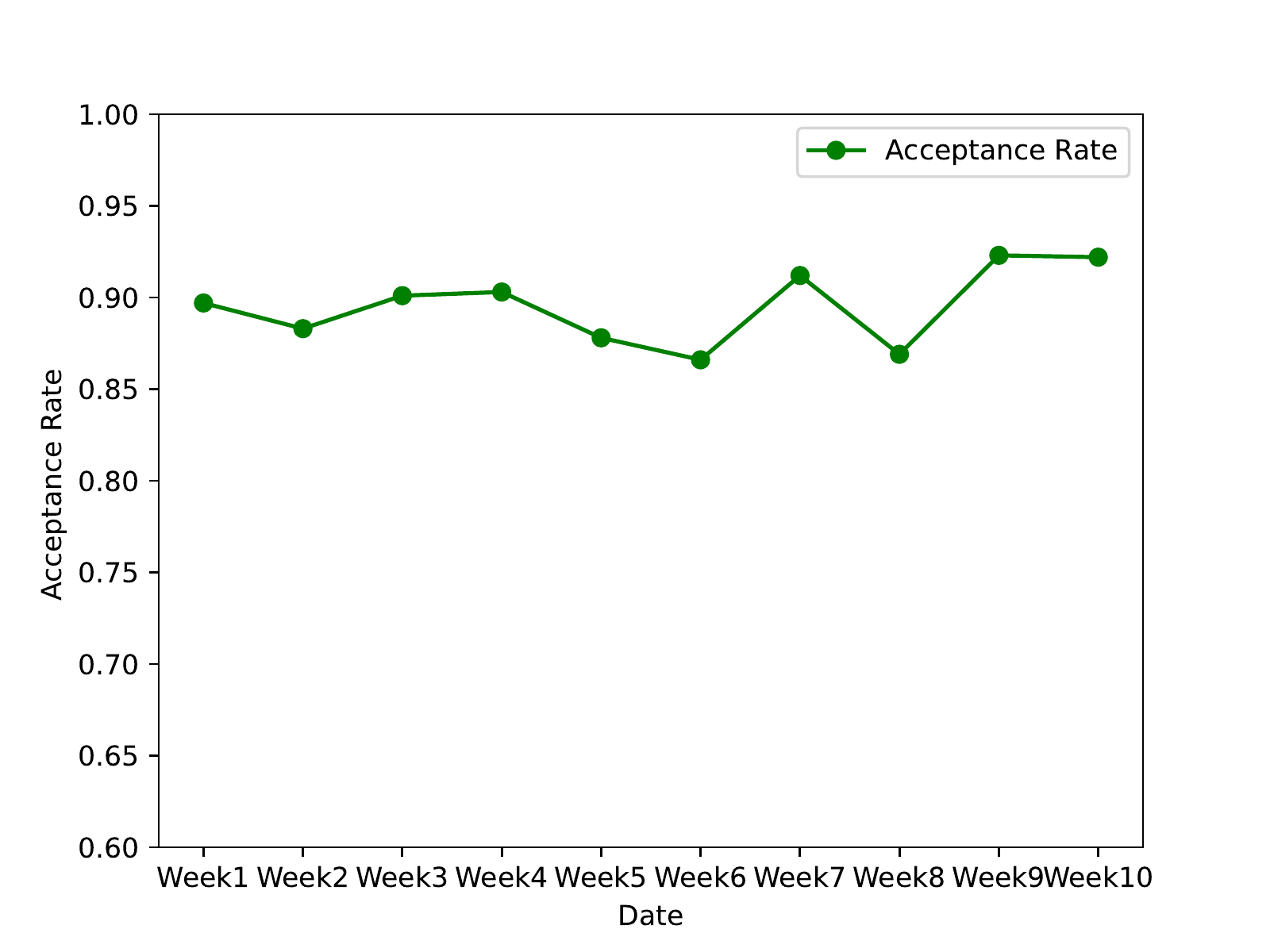}
		\caption{\small Search-based Evaluation}
		\label{search_acceptance_rate}
	\end{minipage}
	\begin{minipage}[t]{0.32\linewidth}
		\centering
		\includegraphics[width=0.88\linewidth]{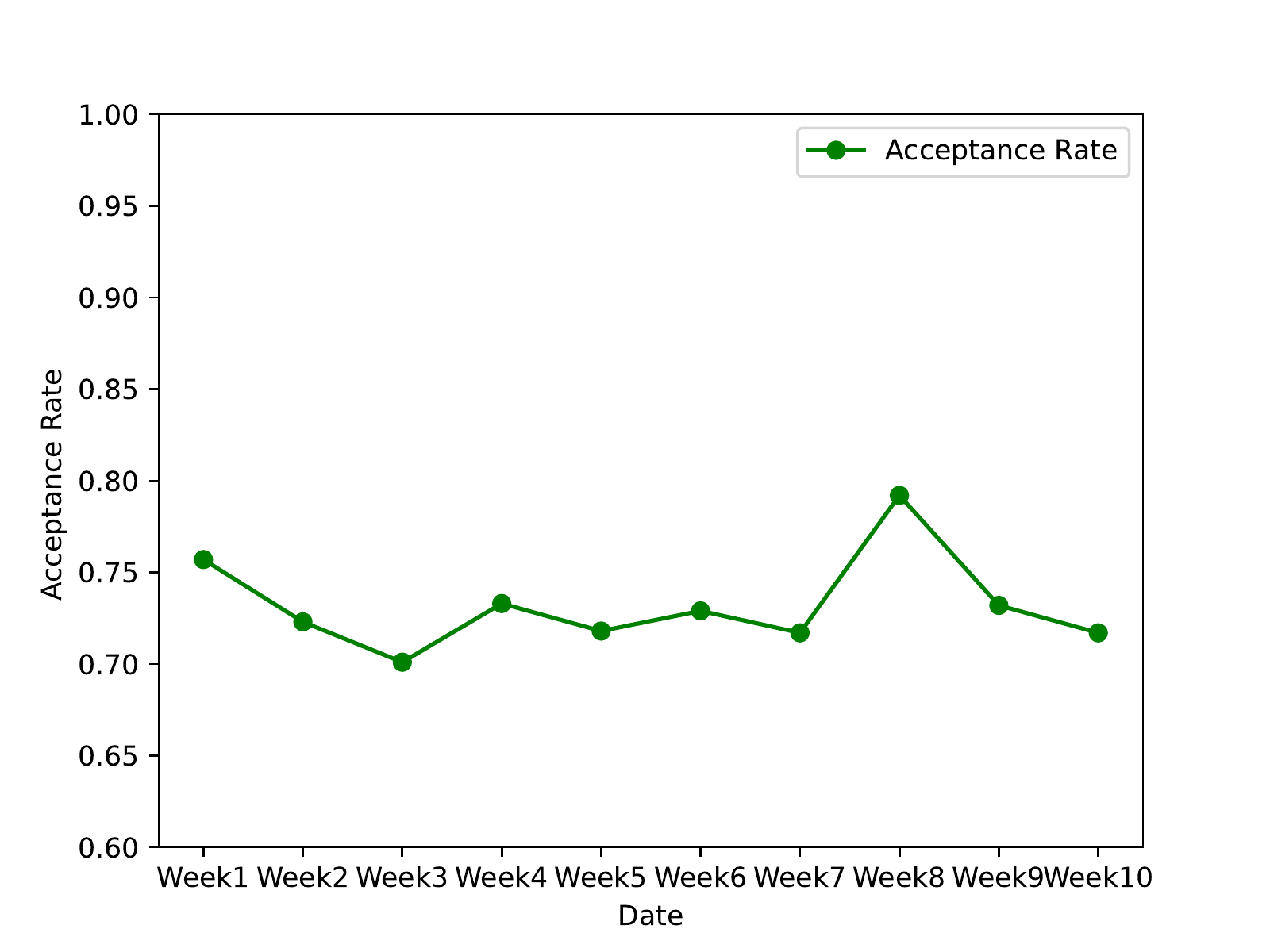}
		\caption{\small Translation-based Evaluation}
		\label{translation_acceptance_rate}
	\end{minipage}
\end{figure*}

\section{Effectiveness Evaluation}
Alibaba Cloud has numerous document development engineers responsible for generating API documents. We deploy gDoc online to observe the acceptance rate of gDoc output results. For example, given a parameter TemplateCode, \textit{gDoc} recommends candidate content. If the documentation engineer selects one of the recommended items as the content of the API documentation, the recommendation is considered valid.
\begin{equation}
	Acceptance~Rate = \frac{Valid~Recommendation}{All~Recommendations}
\end{equation}

\subsection{Overall Evaluation}

As shown in Figure~\ref{overall_acceptance_rate}, we have counted the acceptance rate of gDoc in the last ten weeks, and it is exciting that its acceptance rate has remained above 80\%, which is an excellent testament to the effectiveness of \textit{gDoc}. Up to now, gDoc has been deployed on Alibaba Cloud for more than one year, with a comprehensive acceptance rate of 83.8\%.
\subsection{Search-based Generation Evaluation}

The acceptance rate of the search-based document generation method is maintained at around 90\%. This method considers that there is an intrinsic relationship between APIs, which is often reflected in the fact that if API $\alpha$ and API $\beta$ come from the same producer (or belong to the same category), and $\alpha$ and API $\beta$ contain parameters with the same name, then the two parameters likely to have the same meaning.
\subsection{Translation-based Generation Evaluation}
Unlike the previous integration of API documentation relying on the existing web knowledge, the translation-based method directly produces information and can generate documentation for any API. Although the acceptance rate of translation-based document generation methods is lower than that of search-based document generation methods, the overall acceptance rate is still high. The acceptance rate of 70\%+ is acceptable proof that translation-based methods can be applied to API documentation generation.
\section*{CONCLUSIONS}
In this research, we presented an automatic API documentation generation method \textit{gDoc} to maintain up-to-date documentation with usage examples for an evolving API. Our primary findings provide evidence that \textit{gDoc} can be used as a practical  OpenAPI documentation tool. By leveraging existing API documentation and the Seq2Seq model to document APIs, we have reduced human efforts with automation in the API documentation process while improving the quality of API Documentation. In addition, we discussed techniques to deal with custom content flexible API elements and including the API's documentation with every life-cycle step to establish a quick feedback loop.

\section*{Acknowledgments}

This work was supported by Alibaba Group through Alibaba Innovative Research Program.
\bibliographystyle{ACM-Reference-Format} 
\bibliography{reference}

\end{document}